\newcommand{\degree}{\ensuremath{^\circ}}
\newcommand{\mcdot}{\!\cdot\!}

\newcommand{\bra}[1]{\langle #1|}
\newcommand{\ket}[1]{|#1\rangle}

\newcommand{\eq}[1]{Eq.~\eqref{#1}}

\def\eps{\varepsilon}

\newcommand{\be}{\begin{equation}}
\newcommand{\ee}{\end{equation}}

\def\OMIT#1{{}}

\newcommand{\vc}[1]{{\bf{#1}}}

\newcommand{\ben}{\begin{eqnarray}}
\newcommand{\een}{\end{eqnarray}}

\newcommand{\bef}{\begin{figure}[htb]\centering}
\newcommand{\eef}{\end{figure}}

\documentclass{ws-ijmpcs}

\def\trimmarks{}

\begin{document}

\markboth{Grigory Ovanesyan}
{Angular distributions of higher order splitting functions}

%
\catchline{}{}{}{}{}
%

\title{ANGULAR DISTRIBUTIONS OF HIGHER ORDER \\SPLITTING FUNCTIONS}

\author{GRIGORY OVANESYAN}

\address{Los Alamos National Laboratory, Theoretical Division, Mail Stop B283\\
Los Alamos, NM 87545,
USA\\
ovanesyan@lanl.gov}

\maketitle

\begin{history}
\received{Day Month Year}
\revised{Day Month Year}
\end{history}

\begin{abstract}

We study the angular distributions of the splitting functions for processes for which a parton splits into three partons. Unlike the case of coherent branching, we find that both in vacuum and in the presence of the dense QCD matter, such collinear splitting functions are neither ordered, nor anti-ordered. In the medium-induced splitting functions the angular distributions are broader compared to the similar vacuum distribution, a feature previously noticed from the lowest order medium-induced splitting functions.

\keywords{Parton showers, angular ordering, collinear splitting functions}
\end{abstract}

\ccode{PACS numbers: 12.38.Bx, 12.38.Mh, 24.85.+p}

\section{Introduction}	
In this talk we review our recent results from Ref.~\refcite{Fickinger:2013xwa} on angular distributions of higher order vacuum and medium-induces splitting functions. Such distributions are of interest because parton showers rely on collinear splitting functions. Effects of coherent branchings, see Refs. \refcite{Marchesini:1983bm},\refcite{Marchesini:1987cf}, have been incorporated into parton showers like HERWIG and PYTHIA. Angular-ordered parton shower generates splittings with smaller and smaller values of the opening angles. Such angular-ordered shower has been claimed to include leading infrared logarithms in addition to collinear leading-logarithms that parton shower resum.

The coherent branching result (see Refs. \refcite{Marchesini:1983bm}, \refcite{Marchesini:1987cf} and \refcite{Ellis:1991qj} for review) is that emission of a soft gluon, with the momentum scaling $(\lambda^2,\lambda^2,\lambda^2)$ in light-cone coordinates satisfies the angular ordering condition. This condition states that radiation of the soft emitted gluons is allowed only inside cones between pairs of hard partons in the process, centered on one of them and with the opening angle equal to angular distance between the partons. 

While coherent branchings are valid for long-distance, soft physics observables, another range of distances and energies is of particular interest: the collinear regime. This regime starts after the hard scattering and ends before the ultrasoft recombination regime. To our knowledge, the angular distributions of collinear splitting functions both in vacuum and medium have been first studied in Ref.~\refcite{Fickinger:2013xwa}. Qualitative features of the gluon bremsstrahlung in dense QCD matter, including angular distributions, have  been discussed on the example of a dipole antenna model, see Refs.~\refcite{MehtarTani:2010ma,MehtarTani:2011jw}.
\section{Vacuum splitting $q\rightarrow ggq$}
In this section we calculate the splitting function for a quark to emit two gluons. All such $1\rightarrow 3$ splitting functions have been calculated in Refs.~\refcite{Catani:1999ss}, \refcite{Catani:1998nv}, and we reproduced their result exactly for $q\rightarrow ggq$ splitting. The Feynman graphs contributing for this splitting are depicted in figure \ref{fig:vacuumgraphs}. We work in the light-cone gauge and use SCET as an effective theory for QCD at high energies.
 \bef
\psfig{file=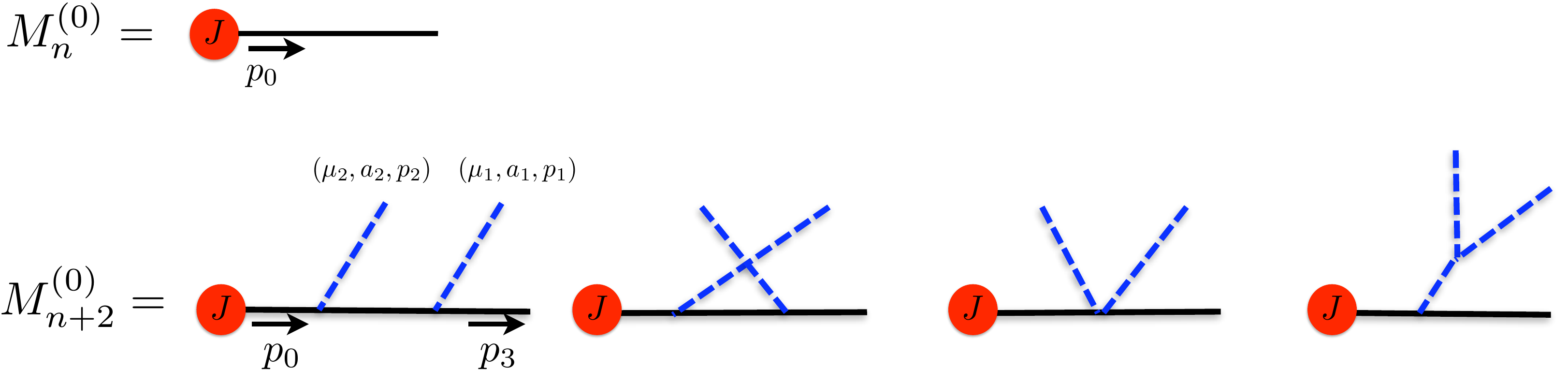, width=3.3in}
\caption{Tree level graphs contributing to vacuum splitting $q\rightarrow ggq$. $J$ represents arbitrary hard amplitude that creates the initial energetic quark.}
\label{fig:vacuumgraphs}
\eef
Using Feynman rules of SCET we get the following expressions for the amplitude before and after the splitting:
\begin{eqnarray}
{\mathcal M}_{n}^{(0)}&=&\bar{\chi}_{n,p_0}J,\\
{\mathcal M}_{n+2}^{(0)}&=&g^2\,\vc{\eps}^{i_1}_{1\perp}\,\vc{\eps}^{i_2}_{2\perp}\,\bar{\chi}_{n,p_3}\,\Gamma_{\text{eff}}^{i_1 i_2}J,\label{eq:mnplus2def}
\end{eqnarray}
where $\chi_{n,p_0}$ is the gauge invariant quark field, $\Gamma_{\text{eff}}$ is found from SCET Feynman rules, see Ref.~\refcite{Fickinger:2013xwa} for more details. The matrix element after the emission factorizes in the following way:
\begin{eqnarray}
\sum\left|{\cal M}_{n+2}^{(0)}\right|^2 &= \frac{4g^4}{s_{123}^2}\langle\hat{P}_{q\rightarrow ggq}\rangle\,\sum\left|{\cal M}_{n}^{(0)}\right|^2\label{eq:Pdef},
\end{eqnarray}
where $\langle\hat{P}_{q\rightarrow ggq}\rangle$ is the spin-averaged splitting function, which is proportional to unit matrix in both color and spin indexes, which we verified by an explicit calculation\footnote{This is true for arbitrary hard process $J$. Similar statement in the presence of dense QCD matter is not correct, see below.}. The splitting function has abelian and non-abelian parts:
\begin{eqnarray}
\langle\hat{P}_{g_1 g_2 q_3}\rangle&=& C_F^2 \langle\hat{P}^{\text{(ab)}}_{g_1 g_2 q_3}\rangle+ C_F C_A \langle\hat{P}^{\text{(nab)}}_{g_1 g_2 q_3}\rangle.
\end{eqnarray}
The abelian part equals to:
\begin{eqnarray}
\langle\hat{P}^{\text{(ab)}}_{g_1 g_2 q_3}\rangle&=&\frac{s_{123}^2}{2s_{13}s_{23}}\frac{z_3(1+z_3^2)}{z_1 z_2}+\frac{s_{123}}{s_{13}}\frac{z_{3}(1-z_1)+(1-z_2)^3}{z_1 z_2}-\frac{s_{23}}{s_{13}}+(1\leftrightarrow 2).\label{finalab}
\end{eqnarray}
The non-abelian part equals to
\begin{eqnarray}
&&\langle\hat{P}^{\text{(nab)}}_{g_1 g_2 q_3}\rangle=\frac{\left[2(z_1 s_{23}-z_2 s_{13})+(z_1-z_2)s_{12}\right]^2}{4(z_1+z_2)^2s_{12}^2}+\frac{1}{4}+\frac{s_{123}^2}{2s_{12}s_{13}}\Bigg(\frac{1+z_3^2}{z_2}+\frac{1+(1-z_2)^2}{1-z_3}\Bigg)\nonumber\\
&&\qquad\qquad-\frac{s_{123}^2}{4s_{13}s_{23}}\frac{z_3(1+z_3^2)}{z_1 z_2}+\frac{s_{123}}{2s_{12}}\Bigg(\frac{z_1(2-2z_1+z_1^2)-z_2(6-6z_2+z_2^2)}{z_2(1-z_3)}\Bigg)\nonumber\\
&&\qquad\qquad+\frac{s_{123}}{2s_{13}}\Bigg(\frac{(1-z_2)^3+z_3^2-z_2}{z_2(1-z_3)}-\frac{z_3(1-z_1)+(1-z_2)^3}{z_1 z_2}\Bigg)+(1\leftrightarrow 2).\label{finalnab}
\end{eqnarray}
In the expressions above the Mandelstam variables are defined in the following way: $s_{ij}=(p_i+p_j)^2$ and $z_i=E_i/(E_1+E_2+E_3)$; the momenta in the $q_0\rightarrow g_1 g_2 q_3$ splitting are defined so, that parton $i$ has four-momentum $p_i$, $i=1,2,3$; $s_{123}=s_{12}+s_{13}+s_{23}$. We refer to final state partons as gluon 1, gluon 2 and quark 3 everywhere below. Expressions \eq{finalab} and \eq{finalnab} are in exact agreements with Refs.~\refcite{Catani:1999ss}, \refcite{Catani:1998nv} for $d=4$ space-time dimensions.

We want to study the angular distributions of this vacuum splitting. In particular we want to explore to which extent, if any, the vacuum splitting $q\rightarrow ggq$ obeys angular ordering? In order to answer this question we need to define the notion of ordering, i.e represent the $1\rightarrow 3$ splitting as a sequence of two $1\rightarrow 2$ splittings. A natural way to define notion of ordering is to take for example, the first parton energy $E_1\ll E_2,E_3$. This can be translated into position space and would mean that the gluon 1 is emitted at longer distances, compared to gluon 2 and quark 3. Thus taking the limit $z_1\rightarrow 0$ allows us to answer the question we asked whether the splitting is ordered or not. One has to keep in mind that in taking this limit we should still have energy of the gluon 1 much bigger than $\Lambda_{\text{QCD}}$, otherwise we would go outside of the collinear regime and our set of approximations becomes invalid.

Taking the described limit $z_1\rightarrow 0$ in the full splitting leads to a simple result:
\begin{equation}
\langle P_{q_0\rightarrow g_1g_2 q_3}\rangle=\frac{4C_F(1-c_{23})}{z_1^2}z_2(1-z_2)\frac{1-z_2+z_2^2/2}{z_2}\left(C_F \left(W_{23}^{[3]}+X_{23}\right)+ C_A\left(W_{23}^{[2]}\right)\right),\\ \label{eq:Pvacsmallz}
\end{equation}
where functions $W_{ij}^{[i]}$ and $X_{ij}$ are the antenna angular ordered and angular anti-ordered functions. For a review see \refcite{Marchesini:1983bm}, \refcite{Marchesini:1987cf}, \refcite{Ellis:1991qj}. These functions satisfy the following equations:
\begin{eqnarray}
&&\int \frac{{\rm d}\phi_{iq}}{2\pi}\,W_{ij}^{[i]}=\frac{1}{1-\cos \theta_{iq}}\,\Theta(\theta_{ij}-\theta_{iq})\label{eq:AO_i},\\
&&\int \frac{{\rm d}\phi_{iq}}{2\pi}\,X_{ij}=\frac{1}{1-\cos \theta_{iq}}\,\Theta(\theta_{iq}-\theta_{ij}),\label{eq:AAO}
\end{eqnarray}
where $\phi_{iq},\theta_{iq}$ are the azimuthal and polar angles between the parton $i$ (either ~2~or~3) and the gluon~1; $\theta_{ij}$ is the angle between partons $i$ and $j$. The meaning of the equations \eq{eq:AO_i} and \eq{eq:AAO} is that the first function emits gluons only in the cone centered at the parton $i$ and with an opening angle $\theta_{ij}$ while the second function emits gluons only outside of two cone centered at $i$ with an opening angle $\theta_{ij}$.

From the properties of functions $W_{23}^{[i]}$ and $X_{23}$ and from \eq{eq:Pvacsmallz} we observe that the abelian piece of the splitting is neither angular ordered, nor anti-ordered (it has both pieces at once), while the non-abelian piece is angular-ordered. This result in a way is very expected. It follows from a simple qualitative argument. When gluon~1 is emitted at an angle much larger than opening angle between gluon~2 and quark~3, it cannot resolve the opening angle between the partons $2,3$ and can see only the initial quark field, which is almost on-shell. Thus in this limit only the abelian radiation is present, while non-abelian vanishes. Thus we obtain again, that the angular distribution is not ordered in the abelian piece but is ordered in the non-abelian one.

In figure \ref{fig:vacuumangular} we present numerical results for the vacuum splitting $q\rightarrow ggq$. The different lines represent abelian, non-abelian and full splittings. The initial decaying quark energy has been set to $E_0=100\text{GeV}$, $z_1=0.03$ and $z_2\approx 2/3$,  $z_3\approx 1/2$; we make sure the energy and momentum are conserved in our kinematics. As a function of the angle between the parton 1 (gluon) and the initial decaying quark we plot the vacuum splitting function. Both abelian and non-abelian parts leak outside of the angular-ordered cones (over angles 40 and 50 with respect to the initial parton). However the non-abelian part falls off as $1/\theta_{01}^4$ (ordered), while the abelian part falls off less steeply $1/\theta_{01}^2$ (not ordered). The reason for which even the angular-ordered part leaks outside of the angular-ordered cone is that in \eq{eq:AO_i} and \eq{eq:AAO} the averaging is over an axis centered at parton $i$ (2 or 3), while in our plot we average over the angle with respect to the initial parton $0$. This numerical result is thus also consistent with our previous qualitative argument.
\bef
\psfig{file=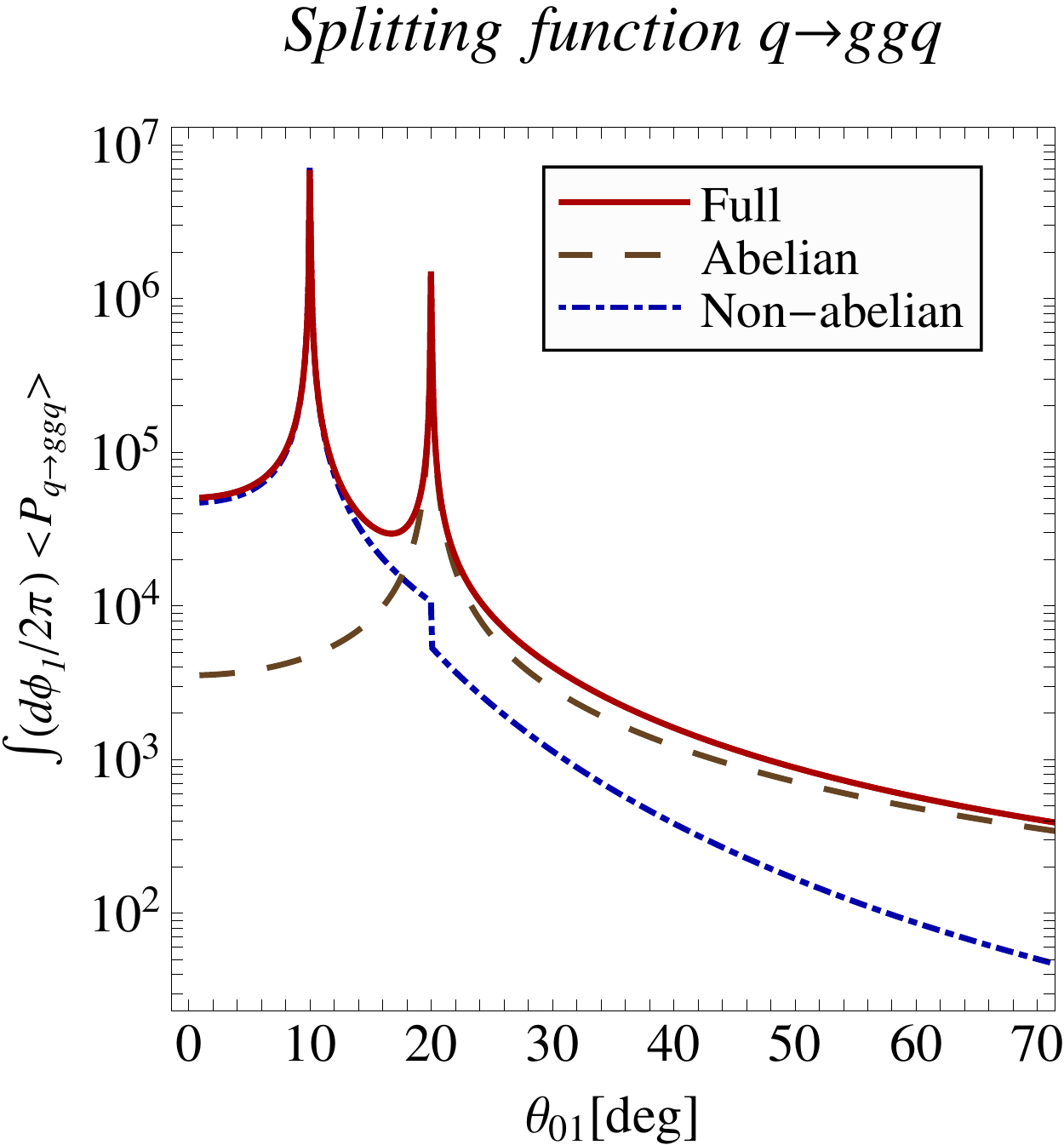, width=2.4in}
\caption{Angular distribution of the splitting $q\rightarrow ggq$ as a function of the angle between the gluon~1 and the initial decaying quark.}
\label{fig:vacuumangular}
\eef
\section{Medium-induced splitting $q\rightarrow ggq$}
In this section we calculate the medium-induced splitting function for the $q\rightarrow ggq$ splitting. We use $\text{SCET}_{\text{G}}$ (see Refs.~\refcite{Idilbi:2008vm,D'Eramo:2010ak,Bauer:2010cc,Ovanesyan:2011xy,Ovanesyan:2011kn,Benzke:2012sz,Ovanesyan:2012fr}) and calculate to first order in the opacity, see for example Ref.~\refcite{Gyulassy:2000er} for general opacity series in the small $x$ approximation.  There are 19 single Born graphs to this order. They are shown in figure~\ref{fig:singleborngraphs}. To each of the columns in the figure we refer to as to topology 1,2,3 or 4. The calculation is simplest in the hybrid gauge, in which collinear gluons are quantized in the light-cone gauge, while the Glauber gluons are quantized in the covariant gauge, see Ref.~\refcite{Ovanesyan:2011xy}. We use this gauge for practical reasons. Each of the single Born graphs takes the following form:
\bef
\psfig{file=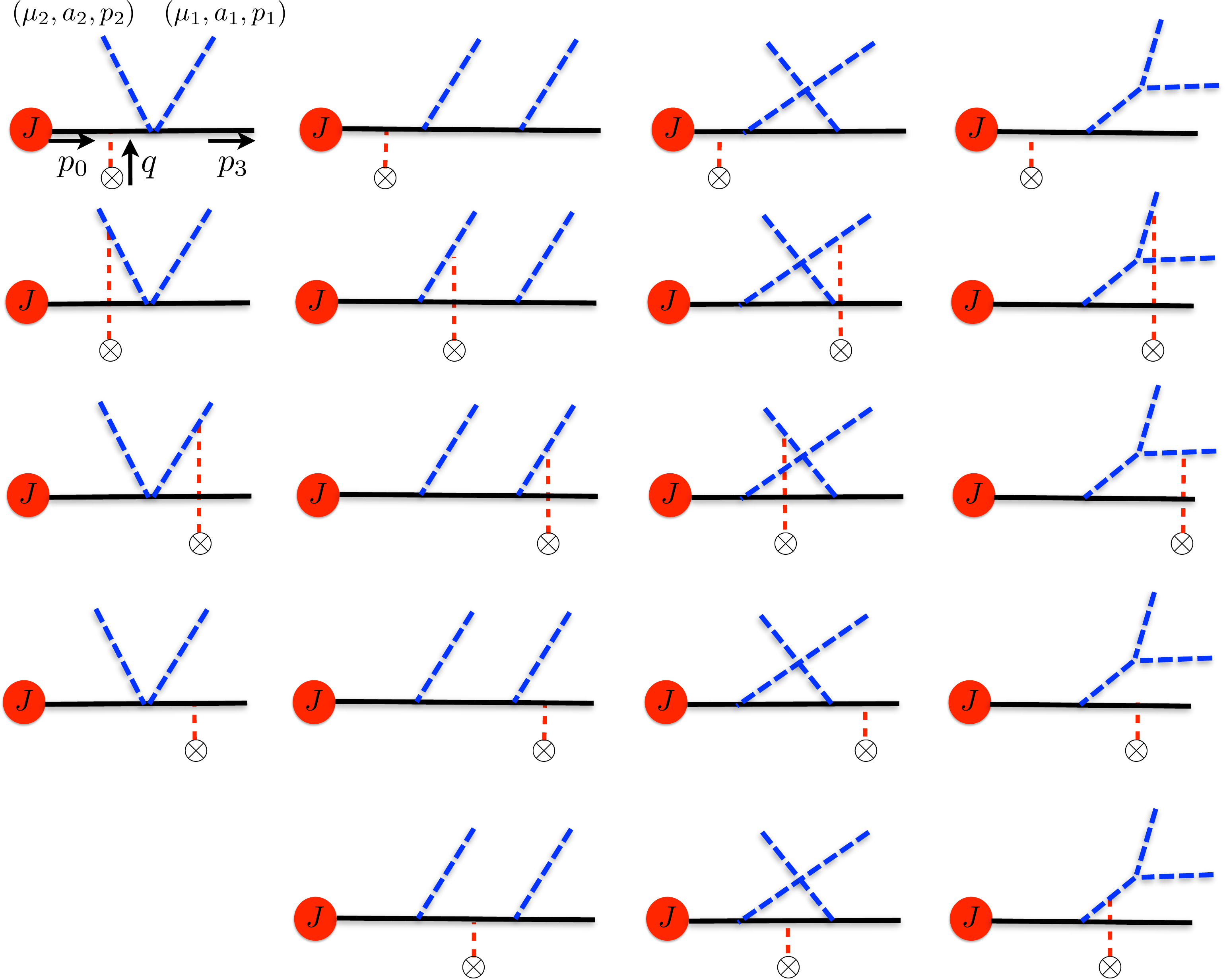, width=3.3in}
\caption{Single Born graphs.}
\label{fig:singleborngraphs}
\eef
\begin{eqnarray}
\mathcal{M}_k^{(1)}=-g^2\,\vc{\eps}_1^{i_1}\,\vc{\eps}_2^{i_2}\,\bar{\chi}_{n,p}\left(\int d{\bf{\Phi}_{\perp}}\,C_k\,\Gamma_k^{i_1 i_2}I_k^{(1)} \right)J,\label{eq:singlebornmaster}
\end{eqnarray}
where integration is over the Glauber gluon transverse momentum, $C_k$ is a number that depends on the topology; $\Gamma_k$ includes the Dirac and color structure and also depends on the topology; finally $I_k^{(1)}$ is the corresponding longitudinal integral for the single Born graph. In Ref.~\refcite{Fickinger:2013xwa} we summarized simple topological rules to determine simply by looking at a given graph the individual pieces of \eq{eq:singlebornmaster} and we refer the reader to there for further details.
\bef
\psfig{file=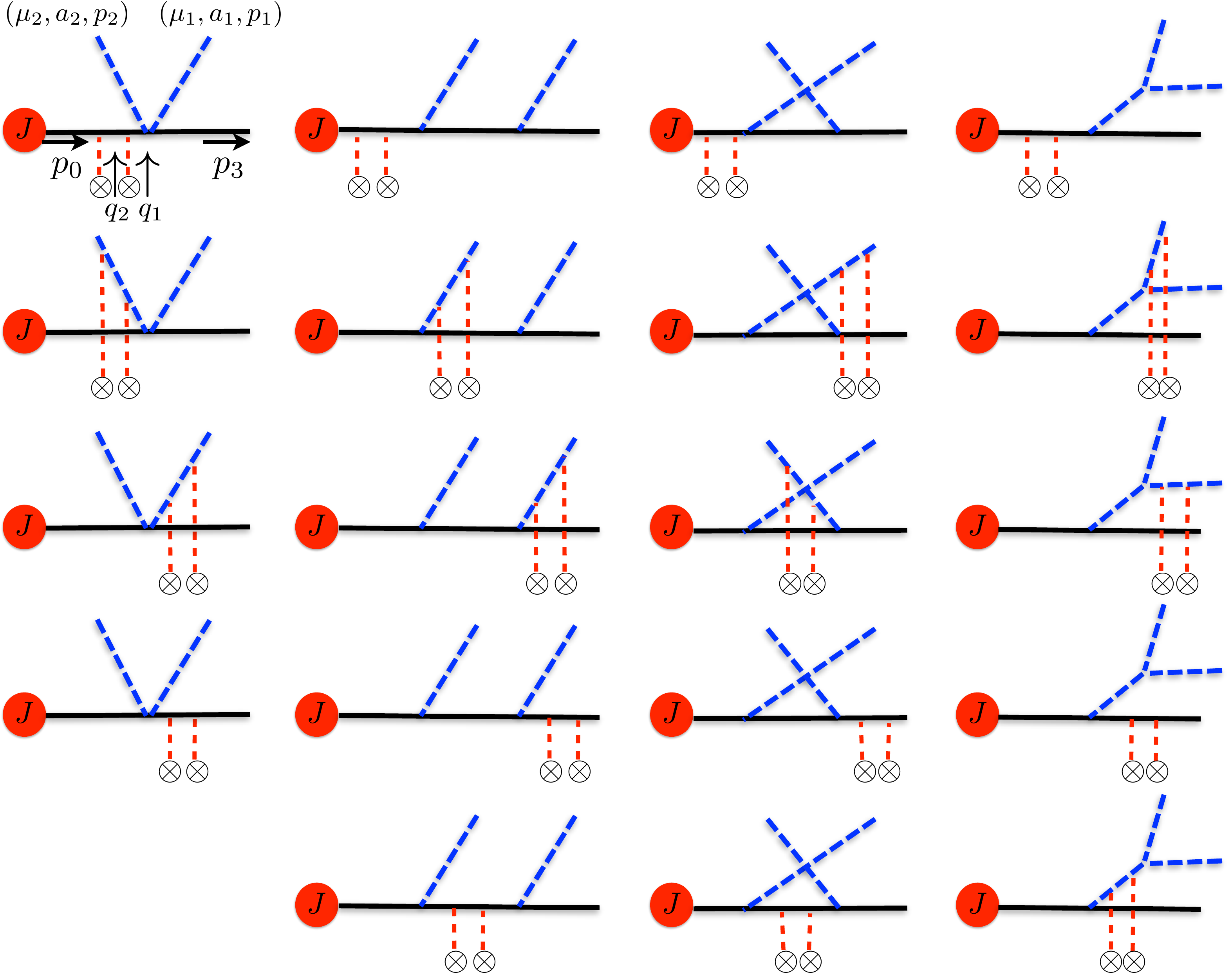, width=3.3in}
\psfig{file=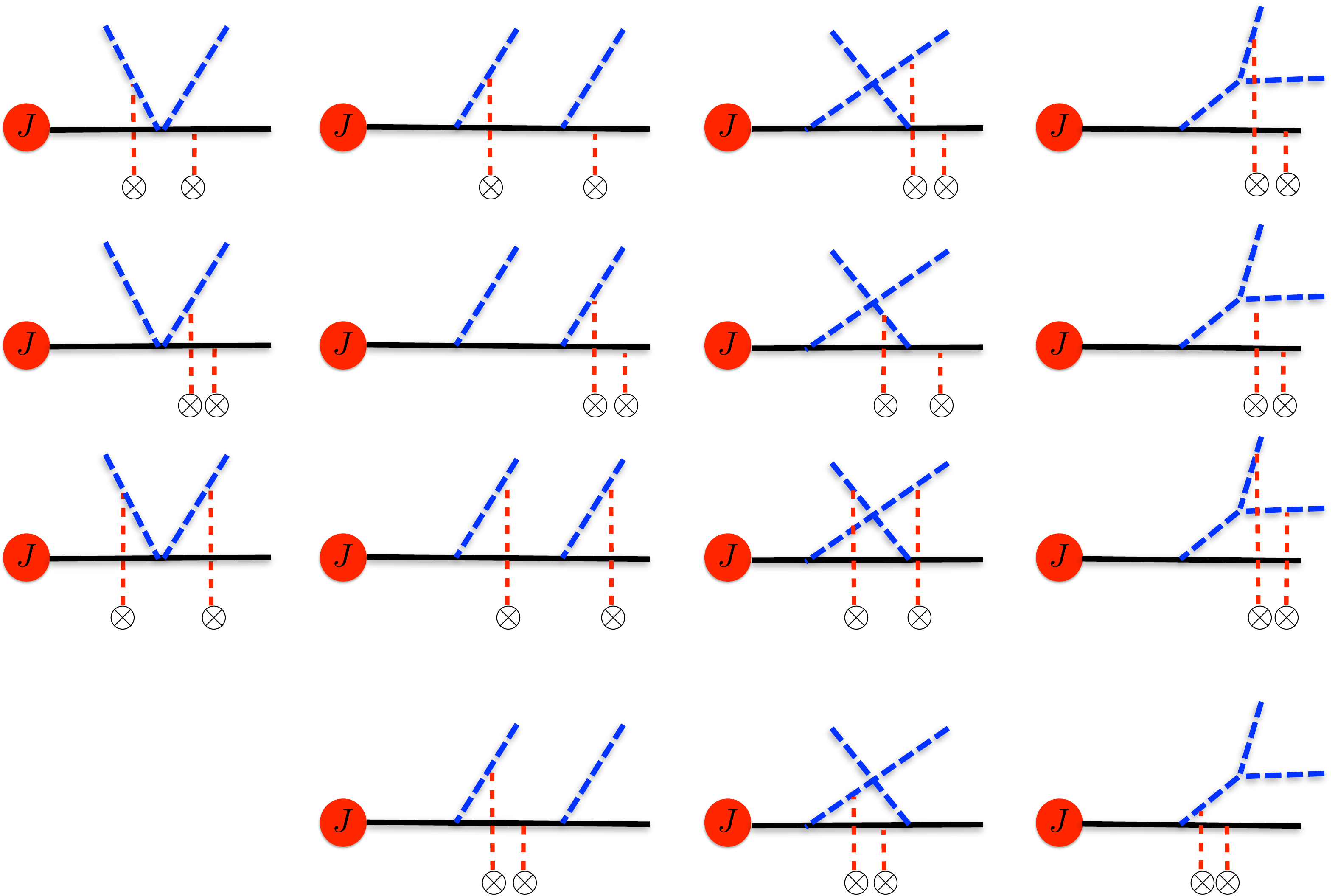, width=3.3in}
\caption{Double Born graphs.}
\label{fig:doubleborngraphs}
\eef

There are 34 total double Born graphs for our calculation and they are all shown in figure \ref{fig:doubleborngraphs}. Similarly to the single Born case, any of these graphs can be written down as: 
\begin{eqnarray}
\mathcal{M}_k^{(2c)}=\,g^2\,\vc{\eps}_1^{i_1}\,\vc{\eps}_2^{i_2}\,\bar{\chi}_{n,p}\left(\int d{\bf{\Phi}}_{1\perp}\,d{\bf{\Phi}}_{2\perp}\,C_k\,\Gamma_k^{i_1 i_2}I_k^{(2c)} \right)J,
\end{eqnarray}
where integration is over the transverse momentum of two Glauber gluons; $C_k$ and $\Gamma_k$ are defined identically to the single Born case; $I_k^{(2c)}$ is the contact limit of the double Born integral. For further details on the calculation and tables for all these entries for each graph we refer the reader to Ref.~\refcite{Fickinger:2013xwa}.

Next we factorize the squared matrix element averaged over the medium, keeping only terms to first order of opacity:
\begin{eqnarray}
\left\langle\sum\left|{\cal M}_{n}^{(0)}+\mathcal{M}^{(1)}+\mathcal{M}^{(2c)}\right|^2\right\rangle_{\vc{q}_{\perp}} = \frac{4g^4}{s_{123}^2}\langle{P}^{(1)}_{q\rightarrow ggq}\rangle\,\sum\left|{\cal M}_{n}^{(0)}\right|^2,\label{eq:mediumfactorization}
\end{eqnarray}
where
\begin{eqnarray}
\langle{P}^{(1)}_{q\rightarrow ggq}\rangle=\frac{2}{N_c}\frac{s_{123}^2}{4}z_3
\int\frac{{\rm d}\Delta z}{\lambda_g(z)}\int{\rm d}^2{\bf{q}}_{\perp}\,
\frac{1}{\sigma_{\text{el}}}\frac{{\rm d}\sigma_{\text{el}}}{{\rm d}^2{\bf{q}}_{\perp}}\left(\rho_1+\rho_{(2c)}\right),\label{eq:Pmediummaster}
\end{eqnarray}
where $\rho_1$ and $\rho_{(2c)}$ are scalar (number) functions that we will present in a moment. However let us pause and mention one assumption that goes into derivation of the equations \eq{eq:mediumfactorization} and \eq{eq:Pmediummaster}. These two equations are derived under the assumption that the hard process $J$ that creates the jet is a pure QCD interaction. For arbitrary process $J$, in particular for jets created by weak interactions the factorization formula \eq{eq:mediumfactorization} contains non-trivial spin correlations. Exactly analogous correlations were found by us in previous work on simpler $1\rightarrow 2$ splitting functions in the medium, see Ref.~\refcite{Ovanesyan:2011xy}. As mentioned in the footnote from the previous section such process dependence is not present in the vacuum case, where for arbitrary process $J$ the splitting function is a singlet in the Dirac space.

In Ref.~\refcite{Fickinger:2013xwa} we provided general formulas for $\rho_1$ and $\rho_{(2c)}$, which ought to be plugged into \eq{eq:Pmediummaster}, thus taking all 19+34 diagrams into account for the medium-induced splitting. In these proceedings we present only reduced results valid only in the limit $z_1\ll z_2, z_3$, which as was explained in the previous section is essential for understanding whether the splitting function is angular-ordered or not. Taking such limit results into the fact that only topologies 2 and 4 are non-zero. Expressions for $\rho_1$ and $\rho_{(2c)}$ become:
\begin{eqnarray}
&&\rho_1\approx 4\left(1-z_2+\frac{z_2^2}{2}\right)
\sum_{k',\,k=1}^{10}\bra{e^{(1)'}_{k'}}\Gamma^{(1)}\ket{e^{(1)'}_{k}}\,\tilde{C}_{k'}^{(1)}\tilde{C}_{k}^{(1)}
\left(\vc{U}_{k'}^{(11)}\,\mcdot\vc{U}_{k}^{(11)}\right)\left(\vc{U}_{k'}^{(12)}\,
\mcdot\vc{U}_{k}^{(12)}\right) \nonumber   \\ 
&&  \hspace{1.9in} \times\text{Re}\,I_{k'}^{(1)*}I_{k}^{(1)},\nonumber\\
&&\rho_{(2c)}\approx 4\left(1-z_2+\frac{z_2^2}{2}\right)\sum_{k'=1,2; k=1,18}\bra{e^{(0)'}_{k'}}
\Gamma^{(2)}\ket{e^{(2)'}_{k}}\,\tilde{C}_{k'}^{(0)}\tilde{C}_{k}^{(2)}\left(\vc{U}_{k'}^{(21)}\,\mcdot\vc{U}_{k}^{(21)}\right)
\left(\vc{U}_{k'}^{(22)}\,\mcdot\vc{U}_{k}^{(22)}\right)\, \nonumber \\  \label{eq:rhomaster}
&& \hspace{2.4in} \times 2\,\text{Re}\,I_{k}^{(2c)},
\end{eqnarray}

where $\tilde{C}_k=R_k\,C_k$ and $R_k=2(1-z_2)$ for topology 2 and $R_k=-2z_2$ for topology~4. For the single Born terms the summation goes over 10 graphs of topologies 2 and 4 in the amplitude and 10 graphs in the complex conjugate of the amplitude. For the double Born terms the summation goes over 18 graphs in the amplitude corresponding to the double Born graphs with topologies 2 and 4, and two graphs in the complex conjugate amplitude corresponding to  vacuum graphs for topologies 2 and 4. The Gram matrices $\Gamma^{(1)}$ and $\Gamma^{(2)}$ are equal to:
\begin{eqnarray}
&&\Gamma^{(1)}=T_R
\left[ \begin{array}{cccccc}
c_1 & c_2 & c_3 & c_2 & c_3 & c_4  \\
c_2 & c_1 & c_2 & c_3 & c_4 & c_3  \\
c_3 & c_2 & c_1 & c_4 & c_3 & c_2  \\
c_2 & c_3 & c_4 & c_1 & c_2 & c_3  \\
c_3 & c_4 & c_3 & c_2 & c_1 & c_2  \\
c_4 & c_3 & c_2 & c_3 & c_2 & c_1 \end{array} \right],\label{gram1}\\
&&\Gamma^{(2)}=T_R\left[
\begin{array}{cccccccccccccccccccccccc}
 c_1 & c_2 & c_3 & c_2 & c_3 & c_4 & c_1 & c_1 & c_2 & c_2 & c_2 & c_3 & c_2 & c_1 & c_1 & c_3 & c_2 & c_2 & c_3 & c_2 & c_1 & c_4 & c_3 & c_2 \\
 c_2 & c_3 & c_4 & c_1 & c_2 & c_3 & c_2 & c_2 & c_3 & c_1 & c_1 & c_2 & c_3 & c_2 & c_2 & c_2 & c_1 & c_1 & c_4 & c_3 & c_2 & c_3 & c_2 & c_1
\end{array}
\right].\nonumber\\ \label{gram2}
\end{eqnarray}
\bef
\psfig{file=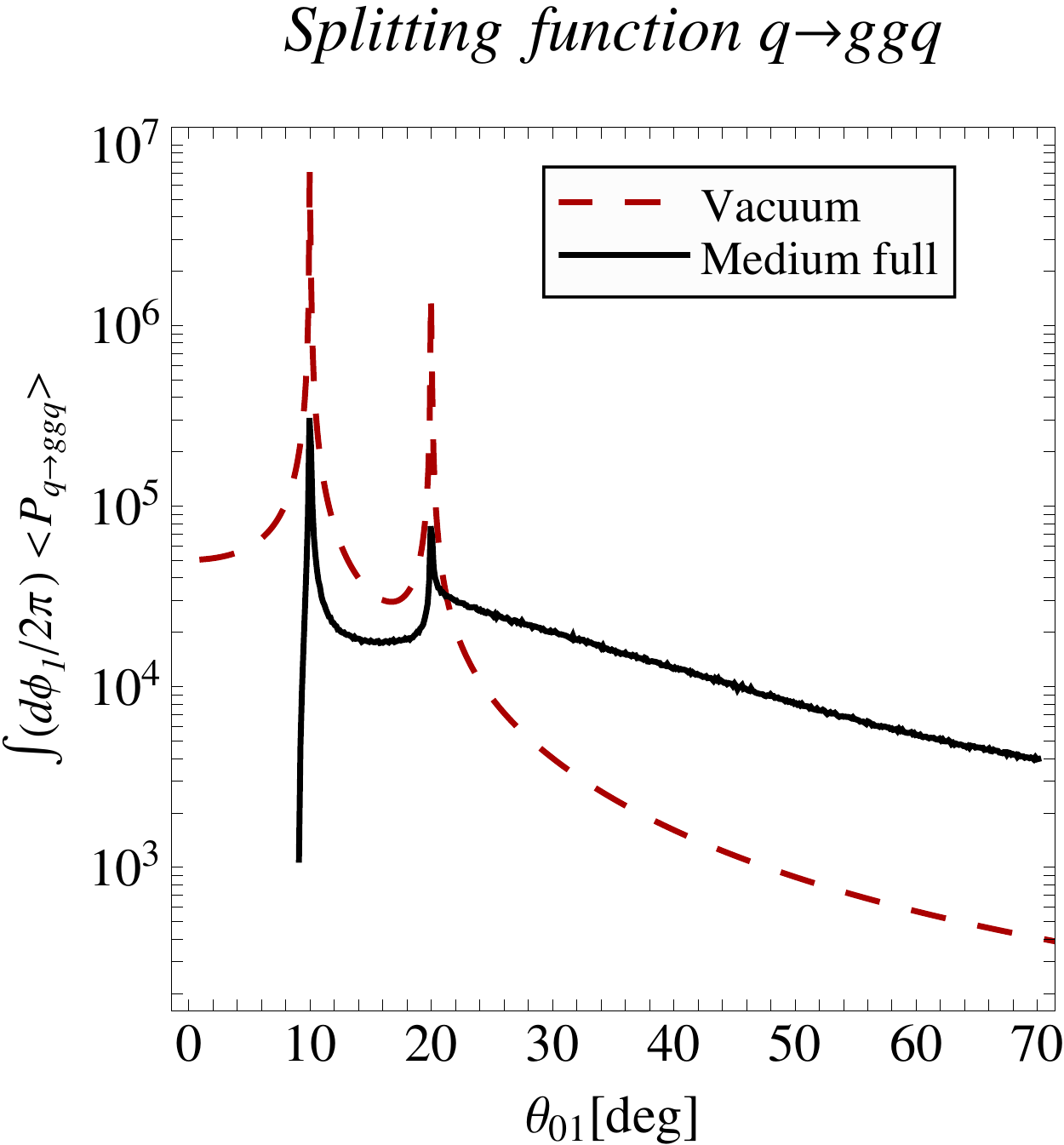, width=2.4in}\,\,\,\,\,\,\,\,\psfig{file=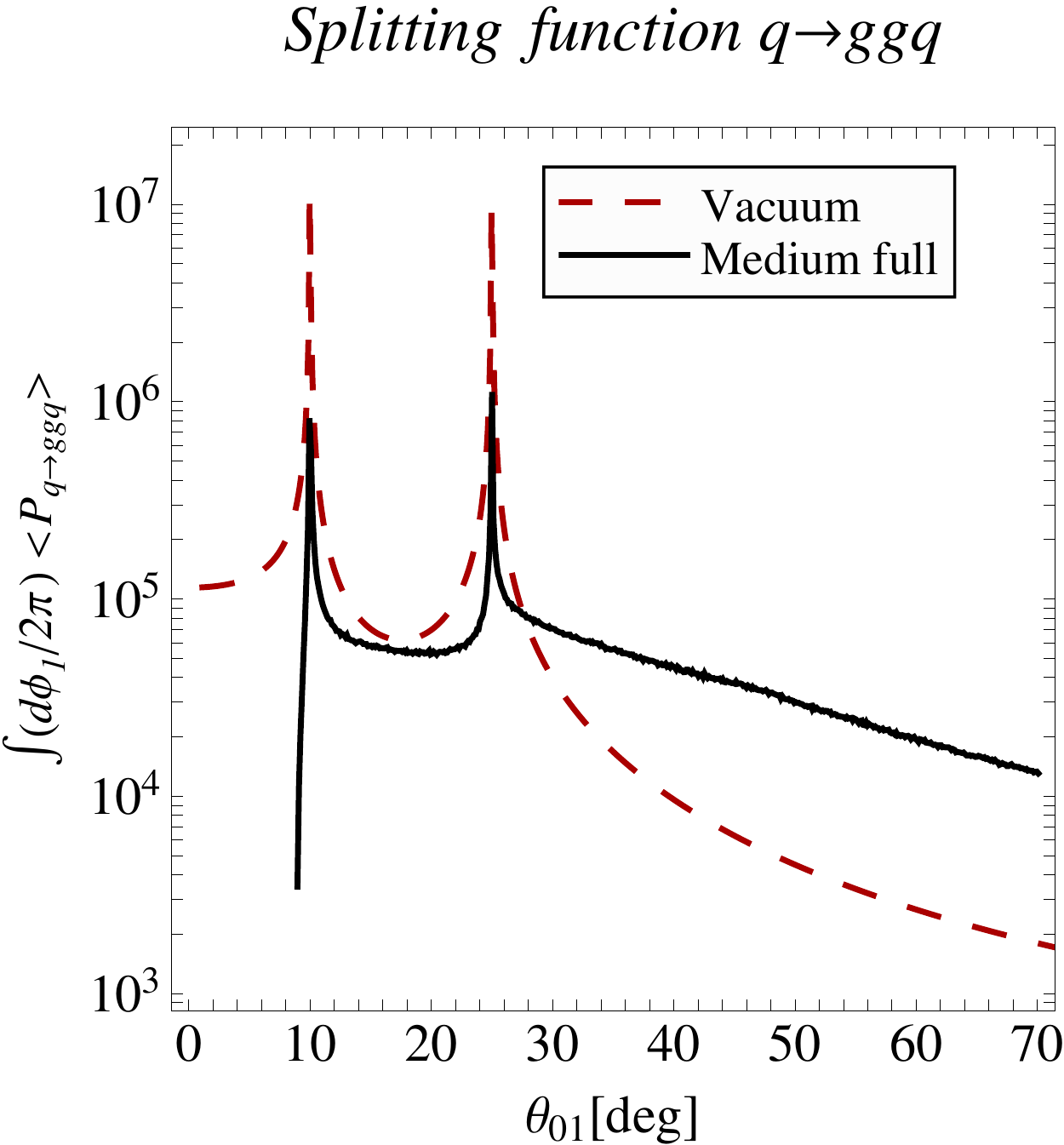, width=2.4 in}
\caption{Angular distribution of medium-induced splitting function for $q\rightarrow ggq$.}
\label{fig:finalnumericsnoordering}
\eef
The color factors $c_1-c_4$ are defined as functions of QCD quadratic Casimirs:
\begin{eqnarray}
c_1&=&C_F^3, \qquad c_2=C_F^2(C_F-C_A/2),\qquad c_3=C_F(C_F-C_A/2)^2,\nonumber\\ 
c_4&=&C_F(C_F-C_A)(C_F-C_A/2)=2c_3-c_2.
\end{eqnarray}
Finally, in equation \eq{eq:rhomaster}, the vectors $\vc{U}_{k}^{(11)}, \vc{U}_{k}^{(12)}$ are first and the second transverse vector in the entry corresponding to the given single Born graph in the third column of the Table 2 in Appendix B of Ref.~\refcite{Fickinger:2013xwa};  $\vc{U}_{k}^{(21)}, \vc{U}_{k}^{(22)}$ are first and the second transverse vector in the entry corresponding to the given double Born graph in the  third column of the Table 3 in Appendix B of Ref.~\refcite{Fickinger:2013xwa}. Coefficients $C_k$ are also given in Tables 2 and 3 of the Ref.~\refcite{Fickinger:2013xwa} and the color operators $e^{(1)'}_{k'}, e^{(2)'}_{k'}$ for each single and double Born graph are given in Appendix B in terms of basis color operators, for which the Gram matrices above were presented.

In figure~\ref{fig:finalnumericsnoordering} we present our numerical results for the medium-induced splitting function for $q\rightarrow ggq$, compared to the vacuum one. For the medium parameters we use $\mu=0.75\,\text{GeV}$, $L=5\,\text{fm}$ and for gluon scattering length $\lambda_g=1\text{\,fm}$, see Ref. \refcite{Ovanesyan:2011xy}. As for the kinematics, we use $E_0=100\text{\,GeV}$ and consider two scenarios. In scenario 1 (left panel of figure~\ref{fig:finalnumericsnoordering}) we use the similar values like in the previous section for vacuum: $E_0=100\,\text{GeV}$, $z_1=0.03$, $z_2=0.643, \theta_{20}=10\degree, \theta_{30}=20\degree$. In scenario 2  (right panel of figure~\ref{fig:finalnumericsnoordering}) we use: $z_1=0.03$, $z_2=0.282, \theta_{20}=25\degree, \theta_{30}=10\degree$.
We present the total medium splitting (solid black curve), and the vacuum splitting (dashed red curve).

From what we already learned from studying the vacuum splitting functions, we already see that the medium-induced splitting function is not angular-ordered, since it has even more radiation than the vacuum one in the tails outside of the angular-ordered cones. While in the forward direction we see less radiation compared to the vacuum case, it is still substantial, so the medium-induced splitting is not anti-ordered.
\section{Conclusions}

We calculated splitting functions for $q\rightarrow ggq$  both in vacuum and in dense QCD matter. In vacuum our result was identical to that of Ref.~\refcite{Catani:1999ss,Catani:1998nv}. We studied the angular distributions of this splitting and found that they are neither angular-ordered, nor angular anti-ordered. The coherent branching result does not hold for the intermediate collinear regime.

In the dense QCD matter the splitting functions are significantly reduced in the forward direction, while in the tail the medium splitting dominates over the vacuum splitting. The same qualitative features have been previously observed in~Ref. \refcite{Vitev:2005yg} for the lowest order $q\rightarrow qg$ splitting function.

The phenomenological applications of our results for parton showers remain to be studied. Of particular interest is finding observables that are more sensitive to the collinear regime and other observables that are more sensitive to softer regime. The next step is to study the effects of turning on and off the angular ordering in the parton shower for the array of such observables.
\section*{Acknowledgments}

This work is supported by DOE Office of Science, Office of Nuclear Physics.


\end{document}